# Identification of asbestos fibres from soil sediments in the Pilsen region of the Czech Republic and the impact of these minerals on the health of the local population

Štěpánka Jansová [1,2,a]

Zdeněk Jansa[1,2], Lucie Nedvědová[1], Ján Minár [1,b]

[1]*University of West Bohemia, Faculty of Ingineering, Departement of Materials and Engineering Metallurgy*
[2] *New Technologies Research Centre, University of West Bohemia, Pilsen, Czech Republic*

[a]Corresponding authors: senjukov@ntc.zcu.cz
[b]jminar@ntc.zcu.cz

**ABSTRACT**

Asbestos is a term commonly used to describe silicate minerals that have a typical fibrous form and crystallize as separable fibers. Due to its extraordinary properties (high tensile strength, resistance to high temperatures, oxidation, corrosion, abrasion, biological degradation, etc.), it has been used for decades mainly in the construction industry. These minerals are usually found as naturally occurring decaying fibres which may be released into the environment as a result of natural processes and anthropogenic activities. Therefore, there is a need to intensify geo-environmental monitoring of the occurrence of natural asbestos on a global scale. However, the study of this material is not only important to clarify the impact of asbestos on public health, but also to know the exact requirements for asbestos replacement materials, which may be man-made mineral and synthetic fibres. At present, the technical but also the environmental reasons for switching to these fibres are more difficult, as asbestos replacement materials are subject to considerable demands - not only technological and economic, but also demands for their biological safety.  The aim of this paper is to summarize the partially current knowledge concerning the extensive problems of asbestos occurrence in general and in the Pilsen region, to search for and possibly determine a suitable methodology for detecting the presence of asbestos in soil sediments based on experimental analyses motivated by analyses in other countries and to determine the exact types of asbestos from one series of samples, and we have been greatly inspired by research in Southern Nevada [19] or Northern Italy [15], where they not only deal with naturally occurring asbestos and its effects on the health of the population, but also incorporate this deliberate mapping of the natural occurrence of asbestos in rocks into laws and decrees of the Ministry of the Environment. This then gives the whole issue the importance it deserves with the great aim of protecting the health of the population.

Samples collected were analyzed by scanning electron microscopy and X-ray diffraction analysis and compared with standards or available literature. Our measurements demonstrated the presence of asbestos in the site sediments and its specific types. The main conclusion of this work is the confirmation of the presence of asbestos in all samples, even its most dangerous types, which can cause very serious diseases. These diseases will be marginally mentioned in our paper. In this context, the mechanism of asbestos-related diseases will be further addressed, which is related to the size and shape of individual fibres, the chemical composition of asbestos types and the links between their basic structural units.

# INTRODUCTION

Asbestos is a naturally occurring fibrous mineral that has been widely used, particularly in the 20th century. These minerals are united by a major common property, which is a fibrous structure. However, in recent decades, considerable research has focused on the strong association between fatal diseases and exposure to asbestos-like minerals present in various environmental matrices. Asbestos is a very dangerous substance which, after years of widespread use in industry and construction, has been identified as a carcinogenic substance with mutagenic effects that poses a risk to the human body and should be treated with caution. When asbestos is disturbed, it releases fibres in the form of dust which can pose a serious health risk. Detailed analytical procedures have so far been established to determine the amount and type of dust particles in the air. However, analyses in soil sediments still leave room for finding the correct and then refining the whole procedure to accurately identify the different types of asbestos. [1][2][3][6]

Asbestos is a subgroup of fibrous minerals - silicates. They are by far the most important group of minerals and, according to educated estimates, make up about 75% of the Earth's crust and, together with quartz (which is structurally close to them), about 95%. Asbestos minerals consist of 40-60% silicon, with the remainder being oxides of iron, magnesium and other metals. A common feature of all asbestos is its fibrous structure, with the length of the fibres being many times their diameter. The most important physical and chemical properties of asbestos fibres are high tensile strength, resistance to acids and alkalis, spinnability, refractoriness, flexibility, elasticity, and poor thermal conductivity. [4][5][10]

Asbestos is divided into 2 main categories: serpentines and amphiboles (Figure 1), whose basic building unit of the silicate structure is the silicon-oxygen tetrahedron $[SiO_4]^{4-}$. The way of combination of tetrahedra in silicate structures is a criterion for their classification. Asbestos minerals that grow in two or three directions instead of breaking into smaller pieces (fragments) are classified as 'non-asbestiform' type asbestos, while these minerals may still have the same chemical formula as the 'asbestiform' type. The term 'asbestiform' type of asbestos refers only to silicates which occur in polyfilament bundles and which consist of extremely flexible fibres of relatively small diameter and long length. These fibres are easily separable from the host matrix or can be cleaved into thinner fibres. A specific characteristic of asbestos is therefore its tendency to form long, thin fibrous structures which tend to split along their entire length. [3][4][5][10][17]

One group of asbestos consists of serpentines, or silicates of magnesium or aluminium or other elements, with an $OH^-$ group. A representative of serpentines is chrysotile, which appears as a white fibre. It is obtained from rocks commonly found throughout the world. Chrysotile fibres are wavy, flexible and tend to form clusters, making it possible to spin and weave them into fabrics. [7][9] The second group of asbestos, the amphiboles, has five members (crocidolite, amosite, tremolite, anthophyllite and actinolite) whose fibres are smooth and have pointed ends (needle-shaped). The most dangerous of the amphibole group is crocodolite, which is often referred to as blue asbestos. [1][2][5]

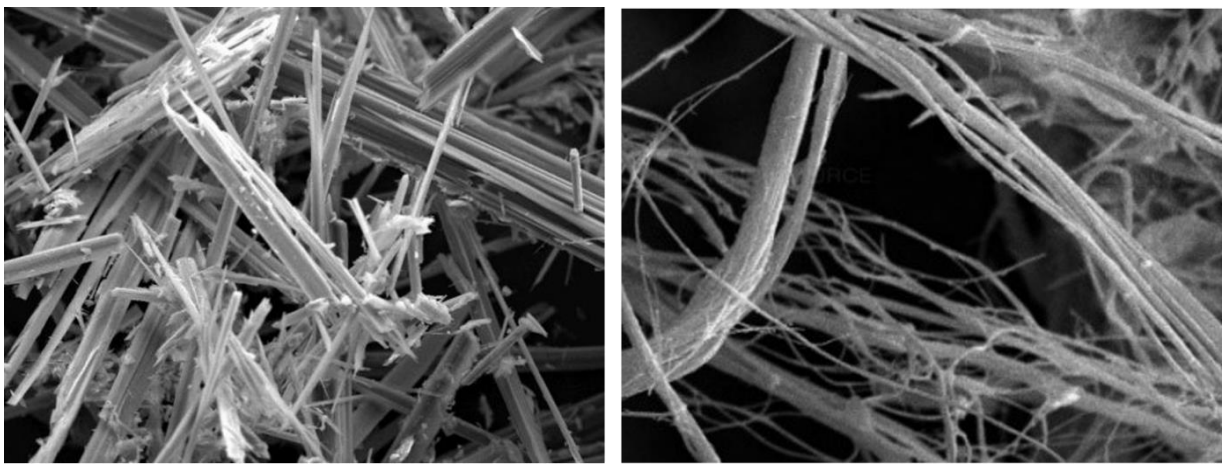

Figure 1. Two basic types of asbestos fibres; electron microscope images:

(a) amphibole type, (b) serpentine type. [1]

# SAMPLE PREPARATION AND EXPERIMENTAL METHODS

Samples for the analyses were taken from four different locations in the same area of the Pilsen region around the Plaska dam, as shown in Figure 2. In view of the increased incidence of cancer in the area, analysis of soil sediments was requested to verify the presence of asbestos and, if confirmed, to determine its exact types. The samples were collected as: sample 1 - sediment of the protruding island, sample 2 - sediment on the shore of the dam, sample 3 - sediment on the road near the dam, sample 4 - sediment from the dog blanket at one of the cottages near the dam. Commonly available tools were used for sampling. The samples were further air-dried naturally, crushed by hand using an agate mortar to a fine-grained fraction, and annealed at temperatures up to 530 °C for four hours. The sediment samples were placed in corundum pellets in the annealing tube of a Carbolite horizontal vacuum tube furnace. The annealing rate of 10°C/min was reduced to 5°C/min after reaching 300°C. After reaching an annealing temperature of 530°C, the temperature was stable for 4 hours. The samples cooled naturally by the action of air. The annealing was done to remove the organic components of the sampled material [16], which facilitated the search for asbestos fibers and associated scanning electron microscope analysis.

The basic elemental analysis and fiber morphology study of the samples was studied using a Quanta 200 thermoemission SEM with an EDS detector in ESEM mode for non-conductive samples without the need for plating. The morphological investigation was carried out by comparing asbestos standards and fibers found by us. Image analysis was carried out using NIS Elements or Fiji (Image J) software to accurately determine the size of the fibres found. All asbestos fibres, without exception, are classified as proven human carcinogens by the International Agency for Research on Cancer (IARC, part of the WHO) and the harmfulness of exposure depends on the type of fibre, its size, the method of processing or on the dose to which the body is exposed. Another analysis was X-ray diffraction, which provided us with an accurate determination of the types of asbestos minerals from the soil sediments collected, based on the known elemental composition of the samples determined by SEM with EDS. Samples were measured using a Panalytical X'Pert PRO automatic powder X-ray diffractometer with geometry Bragg-Brentano with a Cu $K_{\alpha 1}$ radiation source ($\lambda_{K\alpha 1} = 0.154$ nm).

The entire sample preparation, including the collection itself, was carried out under strict security measures (use of resealable and leak-proof bags for storage of the samples collected, protective masks and clothing, cleaning of the collection instruments after each individual collection, respiratory mask with the required level of filters...etc.). Asbestos is a proven human carcinogen and should be handled with great care. For this reason, the need to find a correct and unique procedure for the preparation of soil samples and subsequent analysis is all the more important because working with such a hazardous material causes very serious diseases, which will be discussed later in this article.

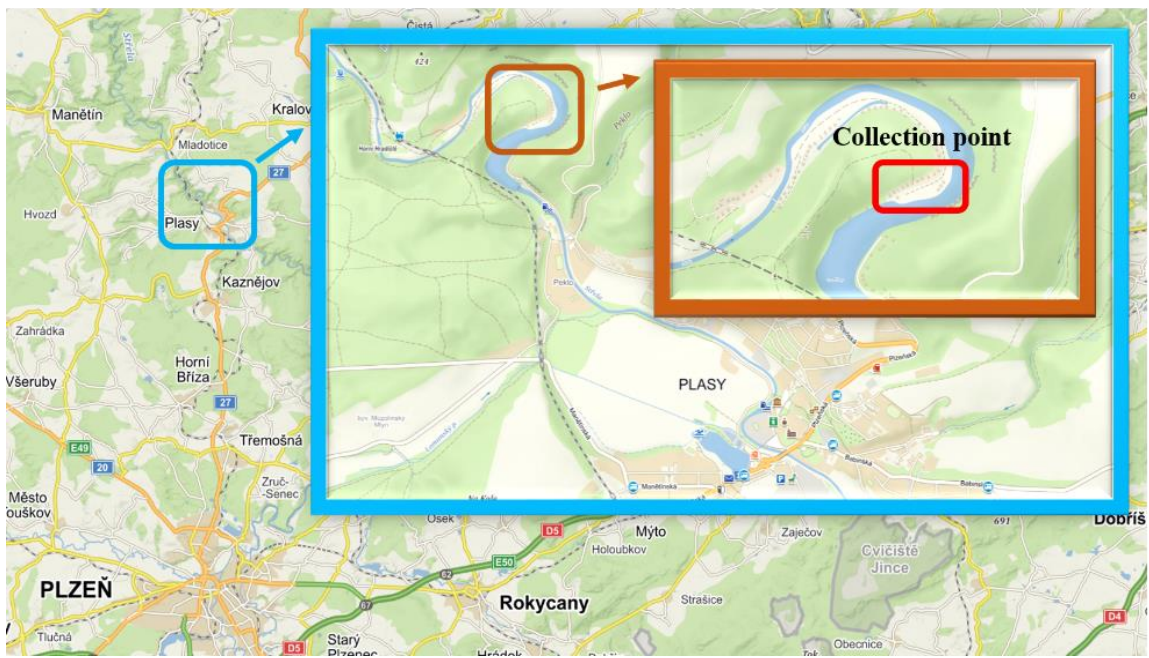

Figure 2. Map showing soil sediment sampling locations. Plaská Dam, Pilsen Region, Czech Republic.

# THE IMPACT OF ASBESTOS FROM A MEDICAL PERSPECTIVE

Asbestos is an occupational and environmental pollutant. Its fibres tend to split along their length, producing very small fibres. If these fibres become airborne, they become flammable. [12] Due to their microscopic size, asbestos fibres can remain in the air for several days after they are released from their basic (either natural or man-made) matrix until they eventually settle in soil or water. [11] Asbestos dust particles have a significant negative impact on human health, especially on the respiratory system. However, the negative effect of asbestos on health is mainly due to mechanical irritation of internal tissues as well as chemical action. It is therefore a combination of irritant, chemical and biological effects on the organism. [13] According to the World Health Organization (WHO), all types of asbestos are classified as carcinogenic and their fibres are considered 'respirable' and hazardous if they are less than 3 micrometres in diameter, more than 5 micrometres in length and have a length to fibre ratio greater than 3:1. [8] No amount of inhaled or ingested asbestos is safe, which means that there is no safe exposure threshold for asbestos.

Health studies report that the long-term presence of asbestos fibres in the human body causes very serious diseases, with the primary organs affected being the lungs, pleura or peritoneum, and in other cases the heart, digestive tract or ovaries. Inhalation of asbestos dust can result in simple dusting of the lungs, called asbestosis, in which healthy lung tissue is replaced by connective tissue. This disease is usually asymptomatic in the long term and is often found quite incidentally on X-ray examination because of the increased deposition of calcium in the connective tissue. Asbestosis, like hyalinosis of the pleura (connective tissue changes in the pleura), can be lived with. But a far more serious disease is bronchogenic lung cancer, a malignant disease more common in smokers than non-smokers. Another of the serious diseases caused by asbestos is pleural or peritoneal mesothelioma (cancer of the pleura or peritoneum). [12][13][14]

Asbestos fibres can be inhaled or ingested. The toxicity of a mineral fibre is closely related to the size and shape of the fibre and its chemical composition. These factors influence whether inhaled fibres penetrate the respiratory tract to the alveolar space and whether the stability of the fibres is such as to cause an increase in their toxicity. At this point, the role is mainly played by macrophages, or natural immunity cells, which play a very important role in the immune response. The macrophage is formed by transformation from monocytes. These are formed in the bone marrow from haemopoietic stem cells and are washed into the bloodstream. The monocytes circulate in the blood for about 8 hours, then enter the tissues and there change into macrophages. Tissue macrophages then show numerous heterogeneities depending on the tissue. The basic function of the macrophage is phagocytosis, a process that ensures the engulfment and processing of foreign, non-functional, dead or diseased cells and other corpuscular material (material size over 100 nanometres). It is the oldest immune process; it can be found in lower animals. [14][20]

Alveolar macrophages phagocytose particles such as dust and microorganisms and remove them from the surface of the alveoli. They also attempt to remove asbestos fibres through phagocytosis. Relatively short fibres appear to be completely encapsulated and removed from the lung compartment by phagosomes (a membrane-bounded vesicle in the cytoplasm that already contains a foreign particle engulfed by the cell during phagocytosis), so that fibres less than 5 μm in length are not retained in the lungs and do not cause chronic inflammation. In contrast, longer fibers are imperfectly phagocytosed by macrophages, leading to "frustrated phagocytosis" (Figure 3) and subsequently remain in the lungs for longer periods of time. Long phagocytosed asbestos fibres are associated with carcinogenesis because they activate the pyrin domain of the NOD-like receptor containing 3 (NLRP3) inflammasome and trigger the production of inflammatory interleukin-1ß (IL-1ß). Damaged and necrotic cells release inflammatory proteins such as high-mobility group box-1 protein (HMGB1), which induce macrophage accumulation and activation of NLRP3, leading to an amplification of the inflammatory response and secretion of tumor necrosis factor-α. Inflammatory cells release reactive oxygen species (ROS) and reactive nitrogen species (RNS) that are capable of causing DNA damage. Thus, phagocytosed asbestos in macrophages causes a mutagenic microenvironment rich in ROS and HMGB1, which increases mutations in mesothelial cells. Chronic inflammation with pleiotropic effect is generated, resulting in malignant transformation. ROS is shown to mediate asbestos-induced DNA damage mutagenesis in human hybrid cells. [14][20]

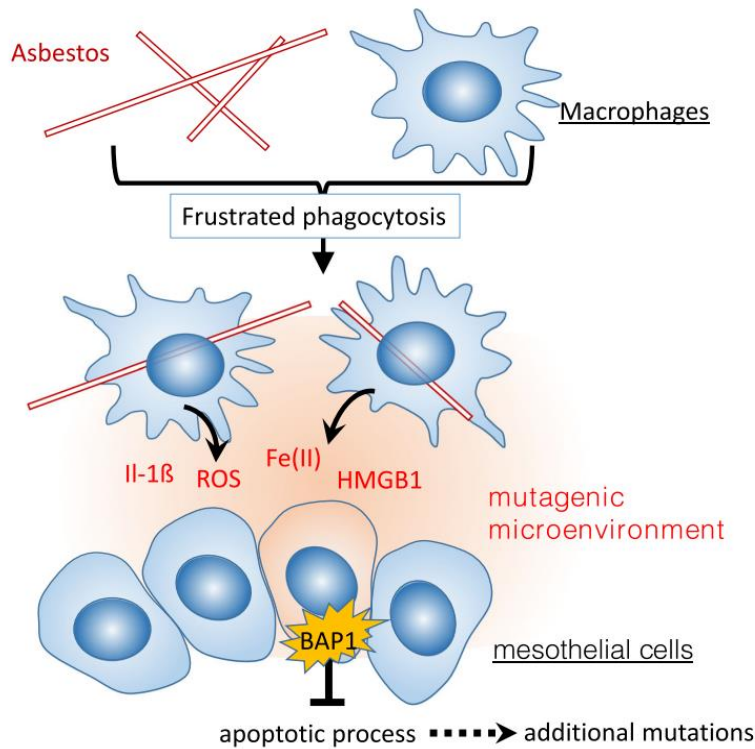

Figure 3: Model of mesothelioma carcinogenesis in an asbestos-induced mutagenic microenvironment. Phagocytosed asbestos induces a mutagenic Fe-rich microenvironment. Mutation of BAP1 contributes to suppression of mesothelial cell death and accumulation of other mutations associated with mesothelioma carcinogenesis. Abbreviations: IL-1ß: Interleukin-1ß; ROS: reactive oxygen species; HMGB1: high mobility group box-1 protein; BAP1: BRCA1-associated protein. [14]

The histological diagnosis of asbestos-related disease toxicity requires the presence of an asbestos fibre 'core' coated with iron-containing materials. Macrophage necrosis occurs repeatedly along with lysosomal cell death, and ferroptosis could create a mutagenic Fe (II)-rich microenvironment. Excess iron is involved in mesothelioma carcinogenesis (e.g., 27% of the total weight of the aphibole is Fe, in serpentine it is 1-3%). Since alveolar macrophages are involved in the transport of substances into the extracellular space or blood, macrophages may be responsible for the transport of asbestos fibers into the pleura and other parts of the human body. For this reason, asbestos can also cause serious diseases in other organs, such as the digestive tract, ovaries, etc. [14][20]

## RESULTS

### Scanning electron microscopy analysis

The fine-grained fraction of the soil samples in the original state without annealing and the annealed fraction at 530°C were attached to the scanning electron microscope sample holder using carbon tape. Fibers were searched for that would correspond in morphology to the mineral fibers of some asbestos group. The fibers found were compared to asbestos type standards and the local elemental composition of these fibers and their surroundings were further measured using an EDS detector.

The above morphological investigation of the first batch of samples showed the presence of fibres whose morphology was consistent with asbestos fibres in all four samples in the pre-annealing condition and in three samples in the post-annealing condition. In the fourth sample, after annealing, this presence was not demonstrated due to the impossibility of analysis because of the lack of material studied. During annealing, most of the fourth sample studied was burned because it contained mostly organic components.

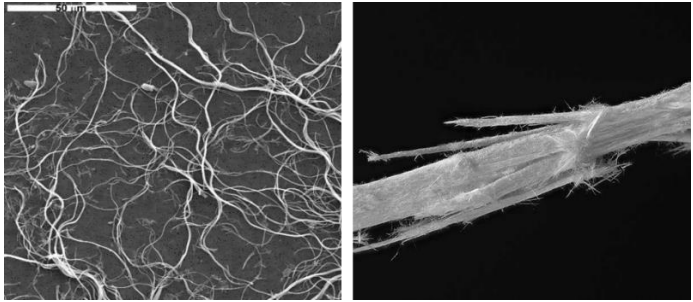

Figure 4: Asbestos Fiber Standards - Serpentine Group [26]

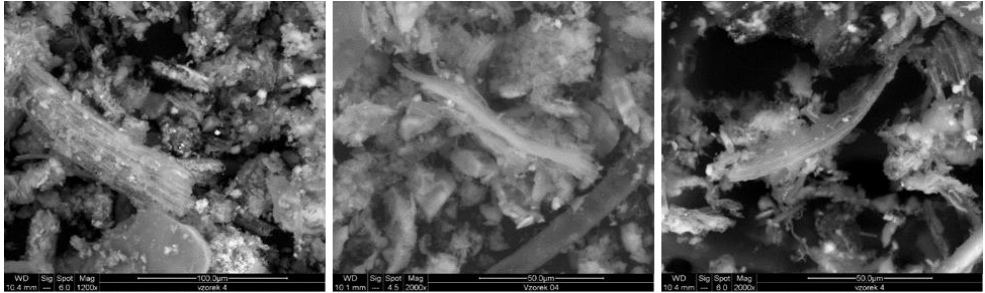

Figure 5: Fibers found, whose morphology according to the comparison corresponded to serpentine-type fibers

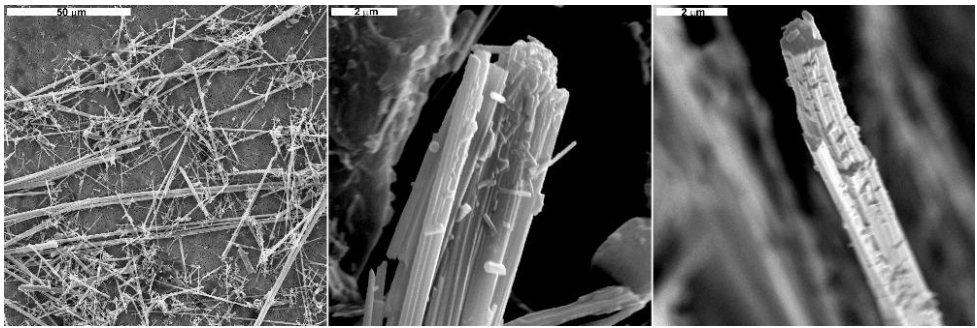

Figure 6: Asbestos fibre standards - amphibole group [27]

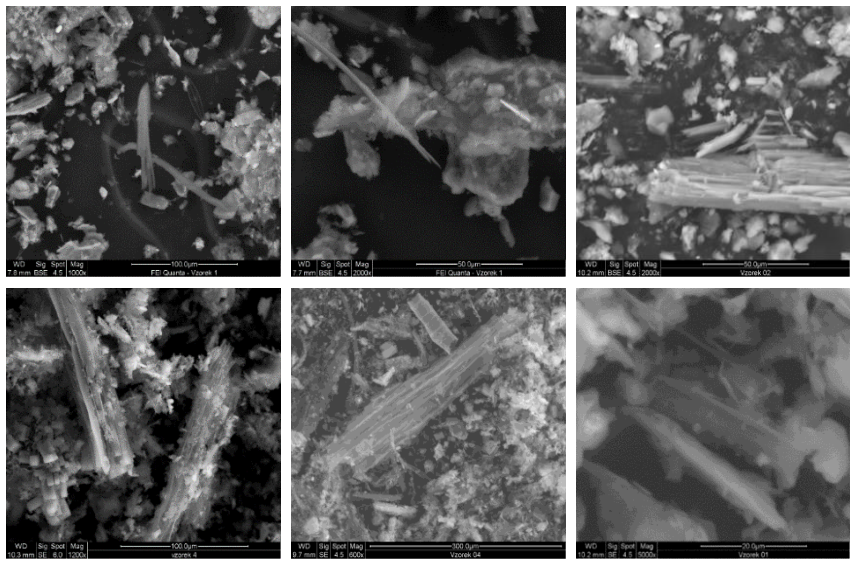

Figure 7: Found fibers whose morphology according to the comparison corresponded to amphibole-type fibers

In the first three samples, smooth fibres with pointed ends were found, corresponding to the amphibole group. In the fourth sample, both smooth fibers from the amphibole group and wavy, flexible fibers from the serpentine group, which tend to clump together, were found. Morphological analysis of these samples was challenging given the low percentage of occurrence of the fibers of interest in the sediment studied overall. Sediments containing a range of mineral and organic constituents made analysis and the possibility of finding asbestos fibres difficult. The images (Figures 5 and 7) show asbestos fibers from all four samples in this series.

## EDS - measurement of local elemental composition

The basic building unit of the silicate structure of asbestos is the silicon-oxygen tetrahedron $[SiO_4]^{4-}$. Chrysotile, as a representative of the first group of asbestos serpentines, is a hydrated magnesium silicate and its stoichiometric chemical composition can be given as $Mg_3Si_2O_5(OH)_4$. However, the geothermal processes that produce chrysotile fibre formations usually involve the simultaneous deposition of various other minerals. It is thus observed that the chemical composition of the fibrous phase is closely related to the composition of the surrounding rock matrix and can thus be variable, as shown in Table 1. [3][4][5][8][21]

The chemical composition of the minerals that make up the second group of asbestos, the amphiboles, reflects the complexity of the environment in which they were formed and can vary considerably in terms of major and trace elements and other contributing factors. The chemical composition of amphibole minerals can be represented as: $A_{0-1}, B_2C_5T_8O_{22}$ (OH, O, F, Cl)$_2$, where A is (Na, K), B is (Na, Ca, Mg, $Fe^{+2}$, Mn, Li), C is (Al, $Fe^{+2}$, $Fe^{+3}$, Ti, Mg, Mn, Cr) and T is (Si, Al), with A, B, C representing cation sites in the crystal structure. From this general representation of the chemical composition of amphiboles it may be deduced that the fibers of amphiboles may be regarded as a series of minerals in which one cation is successively replaced by another, as shown in Table 1. [3][4][5][8][21]

| Mineral | Formula |
|---|---|
| Chrysotile | $Mg_3Si_2O_5(OH)_4$ |
| Amosite | $(Fe_2, Mg)_7Si_8O_{22}(OH)_2$ |
| Crocidolite | $Na_2(Fe, Mg)_3Fe_2Si_8O_{22}(OH)_2$ |
| Antophyllite | $(Mg, Fe_2)_7Si_8O_{22}(OH)_2$ |
| Tremolite | $Ca_2Mg_5Si_8O_{22}(OH)_2$ |
| Aktinolite | $Ca_2(Mg, Fe_2)_5Si_8O_{22}(OH)_2$ |

Table 1: Summary formulas of asbestos compounds indicating the strong variability in the chemical composition of these minerals

The chemical composition was measured repeatedly for all samples, but like the measured chemical composition, the patterns of these minerals are only indicative and idealised given the above information, and therefore the results of the EDS analysis of the soil sediments should be considered only marginally and rather as input information for further analyses. For this reason, also in our experiment the elemental composition analysis was taken as indicative and mainly guided us in the coarse partitioning of serpentines or amphiboles with respect to the presence of some elements that distinguish the different types of asbestos. For example, tremolite and actinolite can be considered as "related" because they differ in composition only in the presence of iron (magnesium in tremolite is partially replaced by divalent iron in the "C" position to form actinolite), sodium indicates the possible presence of crocidolite, and calcium indicates the possible presence of tremolite. Thus, the identification of asbestos fibres from natural sources is not possible on the basis of elemental composition analysis alone, since their composition reflects the natural diversity, variability and constant metamorphosis of minerals. X-ray diffraction analysis was necessary to confirm our hypotheses and results from SEM.

# Image analysis of the SEM images taken with NIS Elements

The found fibers were measured using the NIS Elements. The search was for fibres that met the WHO requirements for the size of 'respirable' fibres. Fibres of this size have been found to be essential for the risk of respiratory cancer or other asbestos-related diseases in humans. Almost all of the fibres found (Table 2) met the WHO requirements when measured for size and were therefore considered "respirable" and therefore highly hazardous to humans.

| sample 1 | | | sample 2 | | | sample 3 | | | sample 4 | | |
| --- | --- | --- | --- | --- | --- | --- | --- | --- | --- | --- | --- |
| lenght [µm] | width [µm] | ratio L/D | lenght [µm] | width [µm] | ratio L/D | lenght [µm] | width [µm] | ratio L/D | lenght [µm] | width [µm] | ratio L/D |
| 16,93 | 0,88 | 19,33 | 8,53 | 0,78 | 11 | 7,81 | 0,85 | 9,19 | 31,33 | 2,8 | 5,4 |
| 22,67 | 2,33 | 9,75 | 9,07 | 1,05 | 8,67 | 11,11 | 1,16 | 9,58 | 12,81 | 2,13 | 6,01 |
| 16,86 | 1,74 | 9,67 | 18,6 | 1,04 | 13,33 | 11,6 | 1,5 | 7,73 | 9,05 | 1,26 | 7,18 |
| 10,22 | 0,88 | 11,67 | 24,42 | 1,74 | 14 | 14,24 | 2,73 | 5,22 | 10,12 | 0,76 | 13,32 |
| 15,91 | 2,34 | 6,81 | 23,26 | 2,33 | 10 | 25,63 | 3,44 | 7,45 | 7,95 | 0,93 | 8,55 |

Table 2. Measured dimensions of some of the fibers found in the first series of the NIS Elements program that meet the WHO criteria for the size of "respirable" fibers.

However, from the point of view of mineralogy and the thorough study of asbestos materials, the infinite cleavability of these minerals must be taken into account. The slightest manipulation of material containing asbestos fibres produces fibres of a size capable of entering the human body and causing serious diseases. Therefore, for the subsequent processing of the next series of samples in our experiment, we eliminate the complex and laborious step of image analysis in the NIS Elements program. In fact, the mere confirmation of the presence of these fibres in the samples will be sufficient to establish the fact that the interference or manipulation of natural deposits (or improper handling of man-made asbestos-containing products) of these minerals will produce fibres dangerous to the human body due to their cleavability.

# X-ray diffraction

As mentioned, X-ray diffraction analysis measurements were chosen to confirm the original hypotheses from the SEM and the presence of asbestos fibres in the samples of this series. Powder X-ray diffraction appears to be a very suitable method for determining the phase composition of the relevant sample together with the possibility of determining the percentage of the relevant asbestos mineral. The determination of the phase composition of asbestos minerals involves a number of problems. Additional phases are present in asbestos-containing soil sediments, the diffraction of which may cause many overlaps in the diffraction maxima of the asbestos phases observed. This results in very similar elemental compositions of the different phases. This problem is solved by progressively refining the measured diffraction record of the sample under investigation. By using databases of standards [24][25]and in conjunction with accurate determination of elemental composition from SEM, the phases present (Table 3) can be accurately determined despite the complexity of the process - thus determining the type of asbestos minerals present in soil sediment samples. [28]

To perform our experiment, the same diffractometer settings were used for all samples to ensure the objectivity of the measurements and the comparability of the results. The measurements were performed on a Panalytical X´Pert Pro powder diffractometer and a copper X-ray tube with a wavelength of K$\alpha$1 = 0.154 nm was used as the X-ray source. A Pixcel ultrafast semiconductor detector was used for data acquisition. The measured data were evaluated in High Score software. All samples were measured at normal atmospheric pressure and at room temperature. Since in all cases the sample was a bulk sample, a standard symmetric geometry with a Bragg-Brentano arrangement was used for the measurements. Due to the hazardous nature of the samples (possible drift of asbestos fibres into the diffractometer compartment or the laboratory), all precautions to prevent drift had to be taken (safe handling of the sample holder, covering the bulk sample until the diffractometer door was closed, use of a disposable protective suit, rubber gloves and respirator). The advantage of using the X-ray diffraction method is that the sample is not mechanically stressed by the machine itself, is not handled in any way during the measurement process and thus cannot be sprayed into the air. However, all related safety precautions must be taken.

The measurement range of all samples was identically chosen in the range of 20 - 85°[2θ]. Within this range, all the strongest diffraction lines of the expected phases are present, both of the possible asbestos and of the accompanying phases such as $SiO_2$. All samples show an enhanced background, especially in the front part of the diffraction record. This is an increase due to the presence of pores and voids in the powder samples, where diffraction on air molecules occurs at these pores and voids. This effect decreases and disappears with increasing angle 2θ. This effect is most pronounced in sample 4, which was taken from a dog blanket. This sample contained the highest proportion of organic matter and therefore could not be properly compacted and the air content of the sample was highest here. On evaluation of the samples, it was found that the main diffraction lines indicating the presence of the asbestos phases of interest were located up to an angle of 40°[2θ], subsequent diffraction lines after this angle belonging only to the $SiO_2$ phase. Therefore, in all diffractograms, the range of angles 20-40°[2θ] is always chosen in the cut-out.

Figures XX to XX show the diffraction records of each sample in turn, before and after firing, with the identified phases marked from 20 to maximum 40°[2θ]. The numerical designation of these phases corresponds to the numerical designation of the identified phases in Table 3. The diffraction line record shows the occurrence of each phase in the samples and it is clear that the presence of asbestos fibres was confirmed by diffraction analysis in all samples of this batch. The results of the X-ray diffraction - phase compositions of each sample are summarised in Table 3. Table 4 then summarises only the asbestos phases present. It is clearly evident from this table that anthophyllite and chrysotile are present in all samples, with the actinolite phase still present in the case of sample 4.

All samples were then subjected to annealing in an oven after initial measurement in their original state (only air-dried to reduce moisture). This procedure was chosen in order to reduce the organic parts present in the samples. These organic phases can cause an increase in the background of the measurements and their presence can cause some of the weaker diffractions of the observed phases to overlap. The effect of annealing is evident in all samples (except for sample 4 - dog blanket, where the organic fraction was so reduced that analysis could not be performed after annealing). In all samples there was a partial reduction of the background present, at the same time there was a reduction of some phases that are sensitive to higher temperatures. The asbestos phases are not affected by this effect, due to their resistance to the temperature used. There is only a reduction or disappearance of some diffractions from inappropriately rotated crystallographic planes.

| Number | Mineral name | Compound name | Reference code | Chemical formula | References |
|---|---|---|---|---|---|
| 1 | quartz a | silicon oxide | 01-089-8935 | $SiO_2$ | [29] |
| 2 |  | magnesium silicate | 01-086-0433 | $Mg_2(Si_2O_6)$ | [31] |
| **3** | **anthophyllite** | **anthophyllite** | **96-901-6382** | **$Mg_{28}Si_{32}O_{96}$** | [30] |
| **4** | **chrysotile** | **chrysotile** | **96-101-0961** | **$Si_{16}Mg_{24}O_{72}$** | [32] |
| 5 | wollastonite 2M | calcium silicate | 01-072-2297 | $CaSiO_3$ | [31] |
| 6 |  | hydrogen silicate | 00-031-0581 | $H_2Si_2O_5$ | [33] |
| 7 | melanophlogite | melanophlogite | 96-901-0372 | $Si_{46}O_{92}C_{6,88}$ | [34] |
| 8 |  | sodium carbonate | 01-086-0300 | $Na_2(CO_3)$ | [31] |
| 9 |  | aluminium silicate | 00-029-0084 | $AlSi_{0,5}O_{2,5}$ | [35] |
| 10 | grossite | calcium aluminium oxide | 00-023-1037 | $CaAl_4O_7$ | [36] |
| 11 | hatrurite | calcium silicate oxide | 01-085-1378 | $Ca_3(SiO_4)O$ | [31] |
| 12 | bernalite | iron hydroxide | 00-046-1436 | $Fe_{+3}(OH)_3$ | [37] |
| **13** | **aktinolite** | **calcium magnesium iron silicate hydroxide** | **00-041-1366** | **$Ca_2(Mg, Fe_{+2})_5Si_8O_{22}(OH)_2$** | [38] |

Table 3. Phases in soil sediment samples identified by the Panalytical X´Pert Pro powder diffractometer [24][25][29][30][31][32][33][34][35][36][37][38]

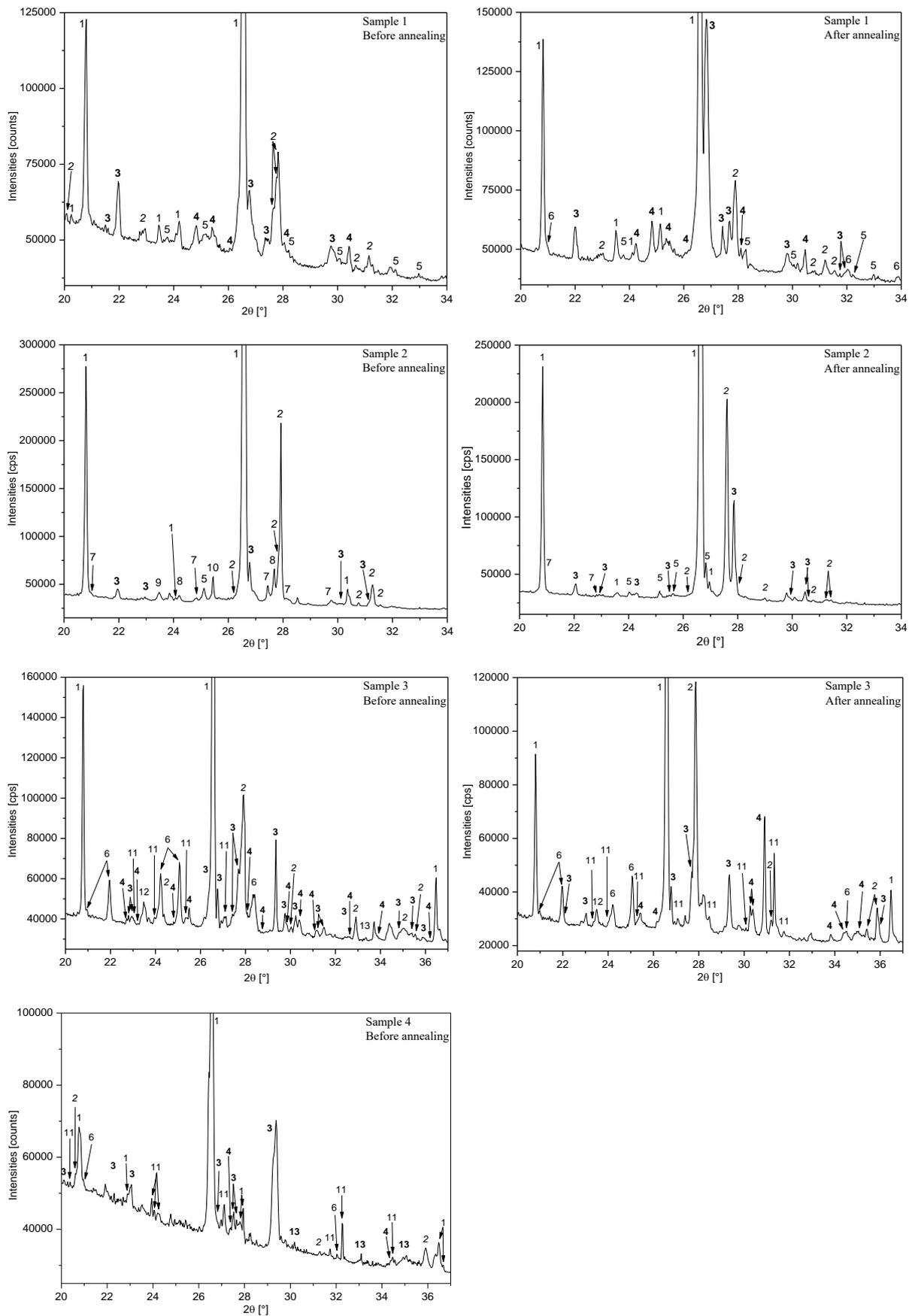

Figure 8a): Diffraction recordings of individual samples before and after firing with the identified phases

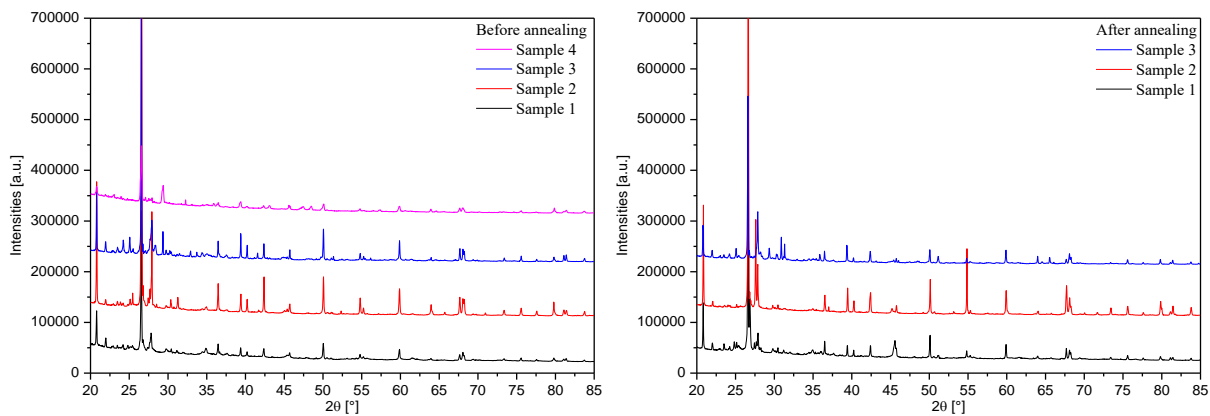

Figure 8b): Comparison of diffraction records of individual samples always before and after firing with the identified phases marked

| Sample | Before / after annealing | Visible Ref. Code | Name | Chemical formula |
|---|---|---|---|---|
| 1 | Before annealing | 96-901-6382 | anthophyllite | $Mg_{28} Si_{32} O_{96}$ |
|  |  | 96-101-0961 | chrysotile | $Si_{16} Mg_{24} O_{72}$ |
|  | After annealing | 96-901-6382 | anthophyllite | $Mg_{28} Si_{32} O_{96}$ |
|  |  | 96-101-0961 | chrysotile | $Si_{16} Mg_{24} O_{72}$ |
| 2 | Before annealing | 96-901-6382 | anthophyllite | $Mg_{28} Si_{32} O_{96}$ |
|  | After annealing |  | anthophyllite | $Mg_{28} Si_{32} O_{96}$ |
| 3 | Before annealing | 96-901-6382 | anthophyllite | $Mg_{28} Si_{32} O_{96}$ |
|  |  | 96-101-0961 | chrysotile | $Si_{16} Mg_{24} O_{72}$ |
|  | After annealing | 96-901-6382 | anthophyllite | $Mg_{28} Si_{32} O_{96}$ |
|  |  | 96-101-0961 | chrysotile | $Si_{16} Mg_{24} O_{72}$ |
| 4 | Before annealing | 96-101-0961 | chrysotile | $Si_{16} Mg_{24} O_{72}$ |
|  |  |  | anthophyllite | $Mg_{28} Si_{32} O_{96}$ |
|  |  |  | actinolite | $Ca_2(Mg, Fe_{+2})_5 Si_8 O_{22}(OH)_2$ |
|  | After annealing | - | - | - |

Table 4. Phases in soil sediment samples identified by the Panalytical X´Pert Pro powder diffractometer [24][29][30][31][32][33][34][35][36][37][38]

## DISCUSSION

From the phase composition results above, it is clear that the presence of asbestos fibres was confirmed in all samples. After completing the results of all analyses, we can say with certainty that chrysotile from the serpentine group is present in the first, third and fourth samples. This is the least dangerous form of asbestos. We can also say that in all samples the presence of anthophyllite from the amphibole group is confirmed and in the fourth sample the presence of actinolite (also from the amphibole group) is confirmed. According to the morphology of the fibres, the third sample most probably also contains crocodolite, but this was not confirmed by XRD. The fourth sample could not be analysed after annealing due to the small sample size. Almost all of the measured fibres from both the pre- and post-annealing samples met the WHO definition and were therefore classified as 'respirable' and highly hazardous. Apart from chrysotile, there is a confirmed occurrence of amphibole types of asbestos, i.e. significantly more hazardous from a health point of view than chrysotile.

The impact of asbestos on human health is well known. The threshold for exposure to asbestos cannot be established medically, as even a single fibre of the 'right' size and composition is enough to cause serious disease. In the initial phase of the research, we searched for fibres in the samples that would meet the WHO conditions with regard to fibre size and respiration, i.e. fibres with a width of less than 3 micrometres, a length of more than 5 micrometres and a length to width ratio of less than 3:1. However, we concluded that this step was unnecessary in view of the infinite cleavability of these minerals and, as a result, the exact type of asbestos found is known. The cleavage of these materials depends on natural influences, mechanical stresses, human intervention, etc.

Previously, fibres shorter than 5 micrometres were not counted in accordance with the standard and were not considered critical. However, their health hazards are not ruled out, and considerable research and studies are under way on the possibility of their riskiness with regard to their chemical composition. The source of contamination of free air with asbestos fibres, especially actinolite in the Pilsen region, has been clearly demonstrated and officially confirmed for many years. For this reason, further detailed investigations of Plzeň and its surroundings are needed due to the significant ecological contamination by asbestos in the air and now in soil sediments, which must be clarified and further investigated in order to achieve better and more objective results and conclusions, to which other specialists from the medical and geological fields are gradually being invited.

## CONCLUSION

Although the presence of asbestos in soil and water can also pose a serious health risk, there has been little research and regulation of asbestos minerals in the past. Several factors influence the pathogenicity of asbestos minerals, the most important of which are the morphology and chemical composition of the fibres. Fibres longer than 10 micrometres cannot be completely absorbed by macrophages and therefore fibre length is still considered to be the main (although not the only) cause of adverse biological effects. [18][23]

Disturbance of asbestos-containing building materials (often through renovation, demolition or reclamation) has been identified as the dominant risk pathway for potential human exposure to asbestos in many areas of the world. However, the significant hazards posed by geologically occurring asbestos, together with other carcinogenic mineral fibres, should also be highlighted. Environmental exposure or dust release during earthmoving activities in areas with geologically occurring asbestos or similar minerals has been less studied; as a result, few management strategies are currently in place. [18][23]

Increased urban development may disturb asbestos-containing rock outcrops or soil containing these and other types of carcinogenic minerals, leading to more exposures, and it is therefore important to establish safe protocols for the identification, extraction, transport and disposal of hazardous soil contaminated with mineral fibres. Thus, all areas near populated areas that may contain asbestiform minerals based on geological studies should be investigated to quantify the risk posed and, if necessary, establish restrictions and procedures to protect construction workers and the general public from exposure. [18][23]

Rocks and soils naturally containing asbestos are therefore also sources of airborne fibres. Concentrations in the environment can vary considerably due to human activities, land use and natural factors. Secondary sources of asbestos, such as soils and waters with natural or anthropic contamination, can release significant amounts of fibres into the air due to this and under certain conditions (i.e. land extraction, slope reclamation, tunnelling, construction activities,...). These secondary sources of asbestos fibres have been and are very little studied and this situation needs to change. The development of techniques for sampling, detection and quantification of asbestos in soil is important for the assessment of sites where contamination is suspected or confirmed. However, there is a lack of standardized analytical methods that provide data on the procedure for collecting these soils and waters for analysis, on a uniform and specific procedure for preparing and processing these samples, and on the analyses of these samples leading to information on the presence of carcinogenic minerals at a given site. [18][23]

In this paper, we studied samples obtained from one locality in the Pilsen region where the presence of asbestos was suspected due to an increased incidence of cancer in the local population. The samples were morphologically examined and compared with standards, analyzed by electron microscopy and X-ray diffraction analysis. This procedure helped us to confirm the presence of several types of asbestos in the region occurring in soil sediments. The presence of asbestos fibers was confirmed in all samples and measured fibers from the samples after annealing met the WHO definition and were therefore determined to be respirable. The main conclusion is that the asbestos found in our four samples is hazardous to the human body and may result in increased cancer rates in the area. It is important to work further with this information to clarify the occurrence of these fibers in the region. It is also necessary to increase the number of samples analyzed for more accurate statistics and more valuable results. It is also necessary to continue to search for and establish a unique procedure for processing soil samples and their analysis for easy and accurate identification of asbestos types without directly endangering laboratory personnel, which will ultimately contribute to a thorough geological mapping in collaboration with other specialists. Previous studies [39] show that the source of air and soil contamination by asbestos fibres from actinolite has been proven and officially confirmed, as it occurs in Proterozoic metamorphosed altered basalts (traditionally called "spilites") in the Pilsen region. It is therefore appropriate to encourage further investigation and mapping of the extent of the area in which the minerals we have confirmed occur: chrysotile, anthophyllite and actinolite, and how to incorporate these findings into the global problem of asbestos and its impact on public health.

## ACKNOWLEDGEMENTS

The result was developed within the specific research project SGS-2021-030 and by the project "Mechanical Engineering of Biological and Bio-inspired Systems", funded as project No. CZ.02.01.01/00/22_008/0004634 by Programme Johannes Amos Commenius, call Excellent Research.

.